\documentclass[12pt]{article}
\usepackage{amsmath}
\usepackage{graphicx,psfrag,epsf}
\usepackage{enumerate}
\usepackage{natbib}
\usepackage{textcomp}
\usepackage[hyphens]{url} 
\usepackage{hyperref}

\newcommand{\blind}{0}

\begin{document}

\def\spacingset#1{\renewcommand{\baselinestretch}%
{#1}\small\normalsize} \spacingset{1}


\if0\blind
{
  \title{\bf A grammar of graphics framework for generalized parallel coordinate
plots}

  \author{
        Yawei Ge \thanks{The authors gratefully acknowledge this research was partially funded by
the 2019 Google Summer of Code} \\
    Department of Statistics, Iowa State University\\
     and \\     Heike Hofmann \\
    Department of Statistics, Iowa State University\\
      }
  \maketitle
} \fi

\if1\blind
{
  \bigskip
  \bigskip
  \bigskip
  \begin{center}
    {\LARGE\bf A grammar of graphics framework for generalized parallel coordinate
plots}
  \end{center}
  \medskip
} \fi

\bigskip
\begin{abstract}
Parallel coordinate plots (PCP) are a useful tool in exploratory data
analysis of high-dimensional numerical data. The use of PCPs is limited
when working with categorical variables or a mix of categorical and
continuous variables. In this paper, we propose generalized parallel
coordinate plots (GPCP) to extend the ability of PCPs from just numeric
variables to dealing seamlessly with a mix of categorical and numeric
variables in a single plot. In this process we find that existing
solutions for categorical values only, such as hammock plots or parsets
become edge cases in the new framework. By focusing on individual
observation rather a marginal frequency we gain additional flexibility.
The resulting approach is implemented in the R package ggpcp.
\end{abstract}

\noindent%
\vfill

\newpage
\spacingset{1.45} 

\newcommand{\pkg}[1]{{\bf #1}}

\newpage

\hypertarget{introduction}{%
\section{Introduction}\label{introduction}}

\begin{figure}
\includegraphics[width=\linewidth]{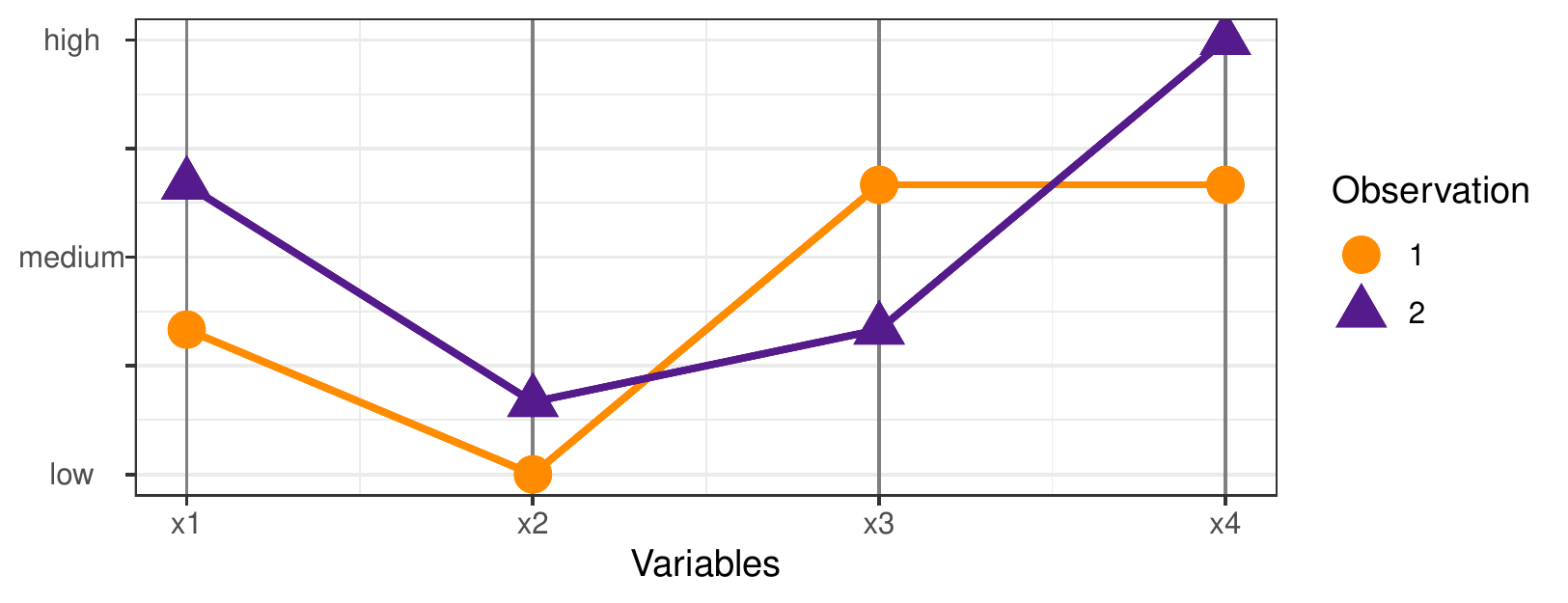} \caption{Sketch of a parallel coordinate plot of two observations in four dimensions. Each dimension is shown as a vertical axis, observations are connected by polylines from one axis to the next. }\label{fig:sketch}
\end{figure}

Few approaches in data visualization exist that are truly
high-dimensional. Most visualizations are (projections of data into) two
or three dimensions enhanced by additional mappings to plot aesthetics,
such as point size and color, or facetting. Parallel coordinate plots
are one of the exceptions: in parallel coordinate plots we can actually
visualize an arbitrary many number of variables to get a visual summary
of a high-dimensional data set. In a parallel coordinate plot each
variable takes the role of a vertical (or parallel) axis (giving the
visualization its name). Multivariate observations are then plotted by
connecting their respective values on each axis across all axes using
polylines (cf.~\autoref{fig:sketch}). For just two variables this switch
from orthogonal axes to parallel axes is equivalent to a switch from the
familiar Euclidean geometry to the Projective Space. In the projective
space, points take the role of lines, while lines are replaced by
points, i.e.~points falling on a line in the Euclidean space correspond
to lines crossing in a single point in the Projective Space. This
duality provides a good basis for interpreting geometric features
observed in a parallel coordinate plot \citep{Inselberg:1985}.
\phantom{@empty}

The origins of parallel coordinate plots date back to the 19th century
and are, depending on the source, either attributed to Maurice
\citet{dOcagne:1885} or Henry \citet{Gannett:1880}. Modern era parallel
coordinate plots go back to \citet{Inselberg:1985} and
\citet{Wegman:1990}. Parallel coordinate plots are used in an
exploratory setting as a way to get a high-level overview of the
marginal distributions involved, to identify outliers in the data and to
find potential clusters of points. In the absence of those, Parallel
Coordinate Plots are often criticized for the amount of clutter they
produce, resembling a game of mikado rather than organized data. This
clutter is sometimes combated by the use of \(\alpha\)-blending
\citep{alpha-blending}, density estimation \citep{density-pcp}, or
edge-bundling parallel coordinate plots \citep{edge-bundling}. For a
detailed overview of these and other techniques see \citet{review}.

However, parallel coordinate plots have some shortcomings. The biggest
challenge comes when working with categorical variables. In current
solutions, levels of categorical variables are transformed to numbers
and variables are then used as if they were numeric. This introduces a
lot of ties into the data, and the resulting parallel coordinate plot
becomes uninformative, as it only shows lines from each level of one
variable to all levels of the next variable. Some versions of parallel
coordinate plots have been specifically developed to deal with
categorical data, e.g.~parallel set plots \citep{Kosara:2006}, Hammock
plots \citep{Schonlau:2003}, and common angle plots
\citep{Hofmann:2013}. These solutions all have in common that they work
with tabularized data and show bands of observations from one
categorical variable to the next. Hammock plots and common angle plots
provide solutions to mitigate the sine-illusion's effects \citep{sine}
on parallel sets plots.

An attempt to combine categorical and numeric variables in a parallel
coordinate plot is introduced in the categorical parallel coordinate
plots of \citet{cpcp}. These plots provide an extension to parallel sets
that allows numeric variables to be included in the plot. Similar to
parallel sets, this approach is also based on marginal frequencies for
the categorical variables. Categorical parallel coordinate plots are the
closest of these variations to our solution, but they are not
implemented in the ggplot2 framework and can therefore not be further
extended.

Various packages in R \citep{R} exist that contain an implementation of
one of the parallel coordinate plots. The function ``parcoord'' in the
MASS package \citep{MASS} make use of the base plot system of R to draw
parallel coordinate plots. The function cpcp in package iplots
implements the parallel coordinate plot \citep{cpcp}. Developments based
on the grammar of graphics \citep{wilkinson:1999} and the ggplot2
\citep{ggplot2} framework are, e.g.~GGally \citep{GGally} or ggparallel
\citep{ggparallel} which provides an implementation of Hammock and
common-angle plots. Those packages based on ggplot2 make use of ggplot2,
but are actually wrapper of existing functions for highly specialized
plots with tens of parameters, which do not allow the full flexibility
of ggplot2 and do not make use of ggplot2's layer framework.

The remainder of the paper is organized as follows: section 2 presents
the conceptual framework of generalized parallel coordinate plots and
general principles informing their construction. Section 3 describes the
connection between generalized parallel coordinates and the grammar of
graphics. Section 4 provides three examples highlighting different
aspects of the use of generalized parallel coordinate plots in an
exploratory setting.

\hypertarget{the-generalized-parallel-coordinate-plot}{%
\section{The Generalized Parallel Coordinate
Plot}\label{the-generalized-parallel-coordinate-plot}}

\begin{figure}
\includegraphics[width=\linewidth]{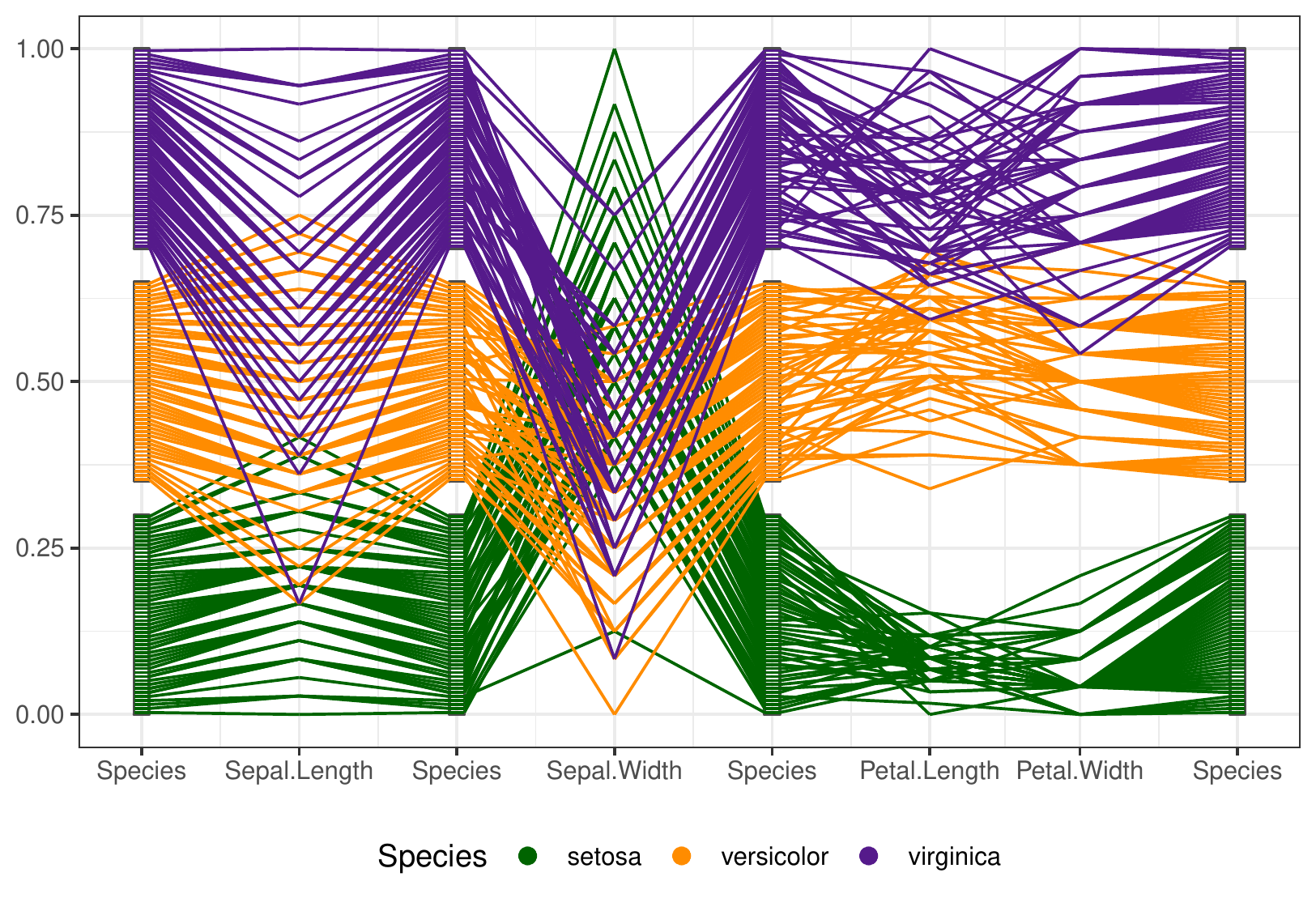} \caption{Generalized Parallel Coordinate Plot of E. Anderson's iris data.  }\label{fig:iris}
\end{figure}

\autoref{fig:iris} shows a first example of a generalized parallel
coordinate plot. Shown are Edgar Anderson's (in)famous iris data
\citep{iris}. Each iris is shown by one polyline. Lines are colored by
species. The species variable is included several times as an axis in
the plot. Sepal widths of irises shows the worst discrimination by
species, their petal widths the best. As can be seen, the categorical
variable species is seamlessly incorporated into the parallel coordinate
plot.

\begin{figure}

{\centering \includegraphics[width=\linewidth]{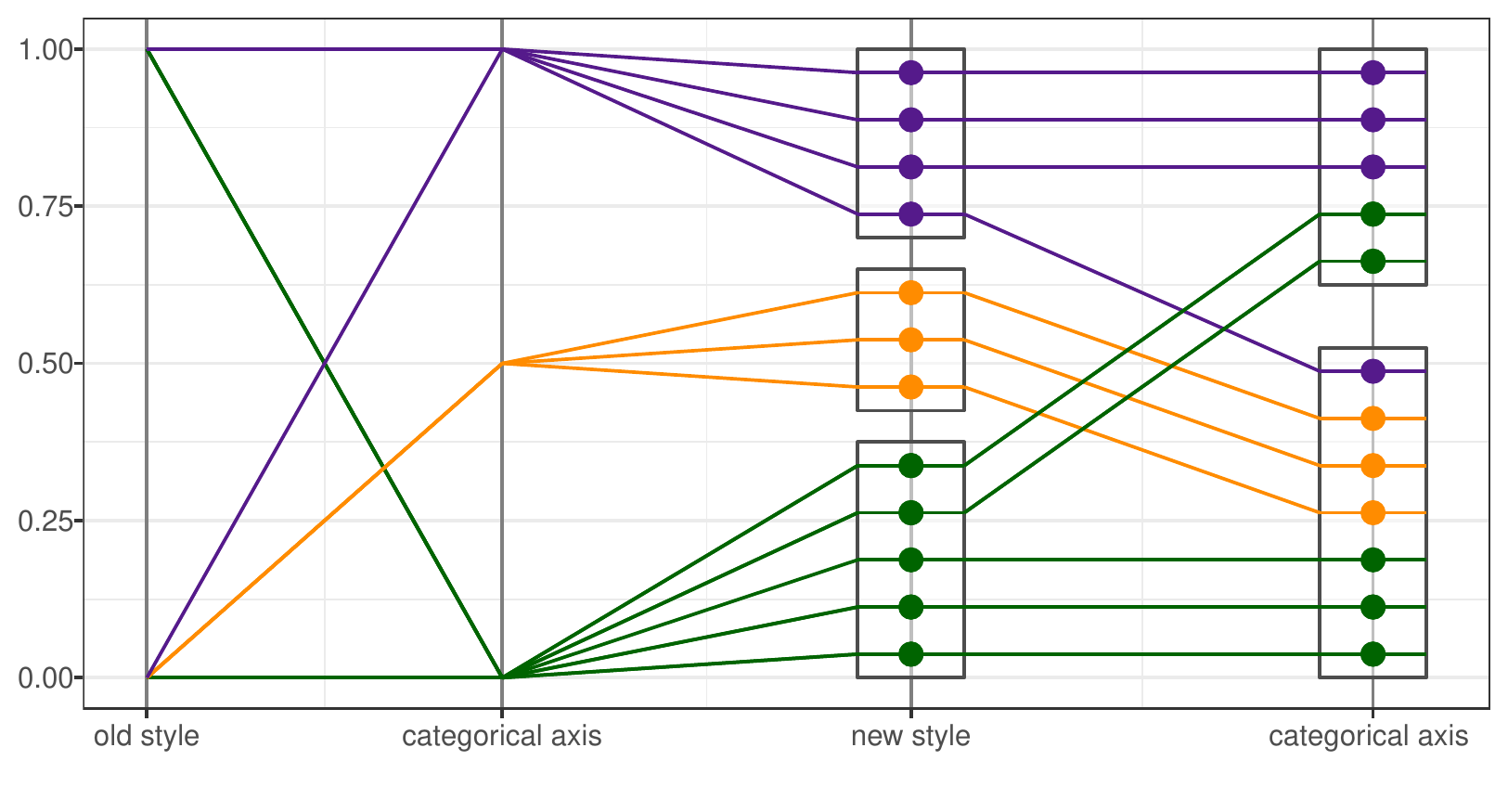} 

}

\caption{Sketch of a parallel coordinate plot with two old-style categorical variables (left) and the same (inverted) categorical axes as implemented in the generalized PCPs (right). }\label{fig:ties}
\end{figure}

The main idea of generalized parallel coordinate plots is that we
integrate categorical variables in a way that allows us to keep track of
individual observations across all variables. This means that we need to
address the inherently existing ties of each level of a categorical
variable. \autoref{fig:ties} shows an implementation of this approach.
The same variables are shown on the left as on the right (in an A-B-B-A
pattern). Instead of using one value for each level, the observations
within each level are spread out uniformly. A rectangle around these
values visually groups all observations of a level together. Some
additional space is inserted between rectangles (for a total of 10\% of
the vertical space) to visually separate the levels. With this
modification we preserve the spirit of parallel coordinate plots by
drawing a trackable polyline across the variables. In comparison, by
using a single point for each level of a categorical variable as on the
left hand side of \autoref{fig:ties}, we end up with a plot that is much
less informative. The most interesting piece of information from the
left hand-side of the plot is that there are no observations in the top
level of the first variable that go into the middle level of the second
variable.

In the remainder of this section we discuss different conceptual aspects
of the construction of generalized parallel coordinate plots before
discussing the specific implementation in the {\bf ggpcp} package.

\hypertarget{breaking-ties}{%
\subsection{Breaking Ties}\label{breaking-ties}}

What might not be apparent at first glance is that the order of the
observations within each level has to be chosen carefully in order to
create a visualization with as little visual clutter as possible. As all
of the observations in each level of a categorical variable share the
same value, the ``values'' on a categorical axis are not determined by
the data, instead they are just positions based on tie-breaking
considerations and therefore provide us with a lot of freedom in
choosing them. This way we are able resolve a lot of potential line
crossings between adjacent axes.

\begin{figure}
\includegraphics[width=\linewidth]{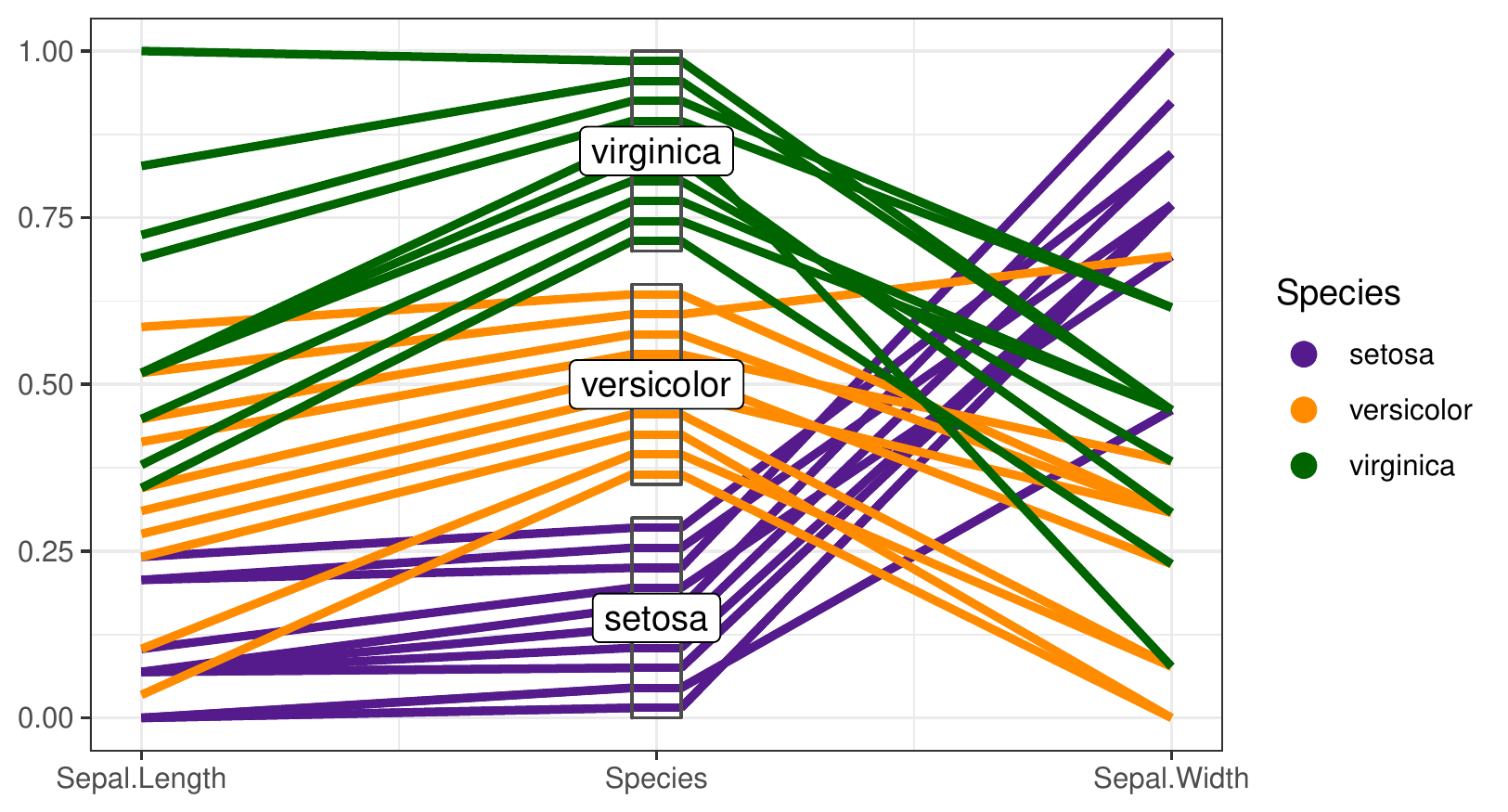} \caption{Numeric-categorical case, values in each level of the categorical variable are sorted according to the values in the numeric variable on the left.}\label{fig:NCN}
\end{figure}

There are four combinations of adjacent variables (N-N, N-C, C-N, and
C-C) to be considered with respect to their tie-breaking behavior in
constructing generalized parallel coordinate plots. While there might be
ties in some numeric variables, we are not changing any of the
established behavior of parallel coordinate plots, i.e.~values any
numeric variable are marked along the axis (after appropriately scaling
the axis) and connected to their respective counterparts on adjacent
axes. When we are dealing with adjacent numeric and categorical
variables (N-C, C-N), we use the values of the numeric variable to
inform the position of the observations within each level of the
categorical variable. Note, however, that it is not possible to sort a
categorical variable with respect to numeric variables on both sides
(N-C-N), shown in \autoref{fig:NCN}. In that situation, our choice is to
always sort the categorical variable according to the numeric variable
on the left hand side first and only regard the values on the right if
there is no numeric variable on the left.

In dealing with adjacent categorical variables, we make use of the basic
idea of parallel set plots \citep{Kosara:2006}, applied to individual
observations rather than based on frequencies, with the aforementioned
advantages. For categorical variables, we extend the construction of tie
breakers to include all adjacent categorical variables. We will call
these sets of categorical variables factor blocks. Within each factor
block the position of observations within each level of a categorical
variable is informed by the joint distribution of a categorical variable
with all of its right-sided categorical neighbors.

\begin{figure}[h]
\includegraphics[width=\linewidth]{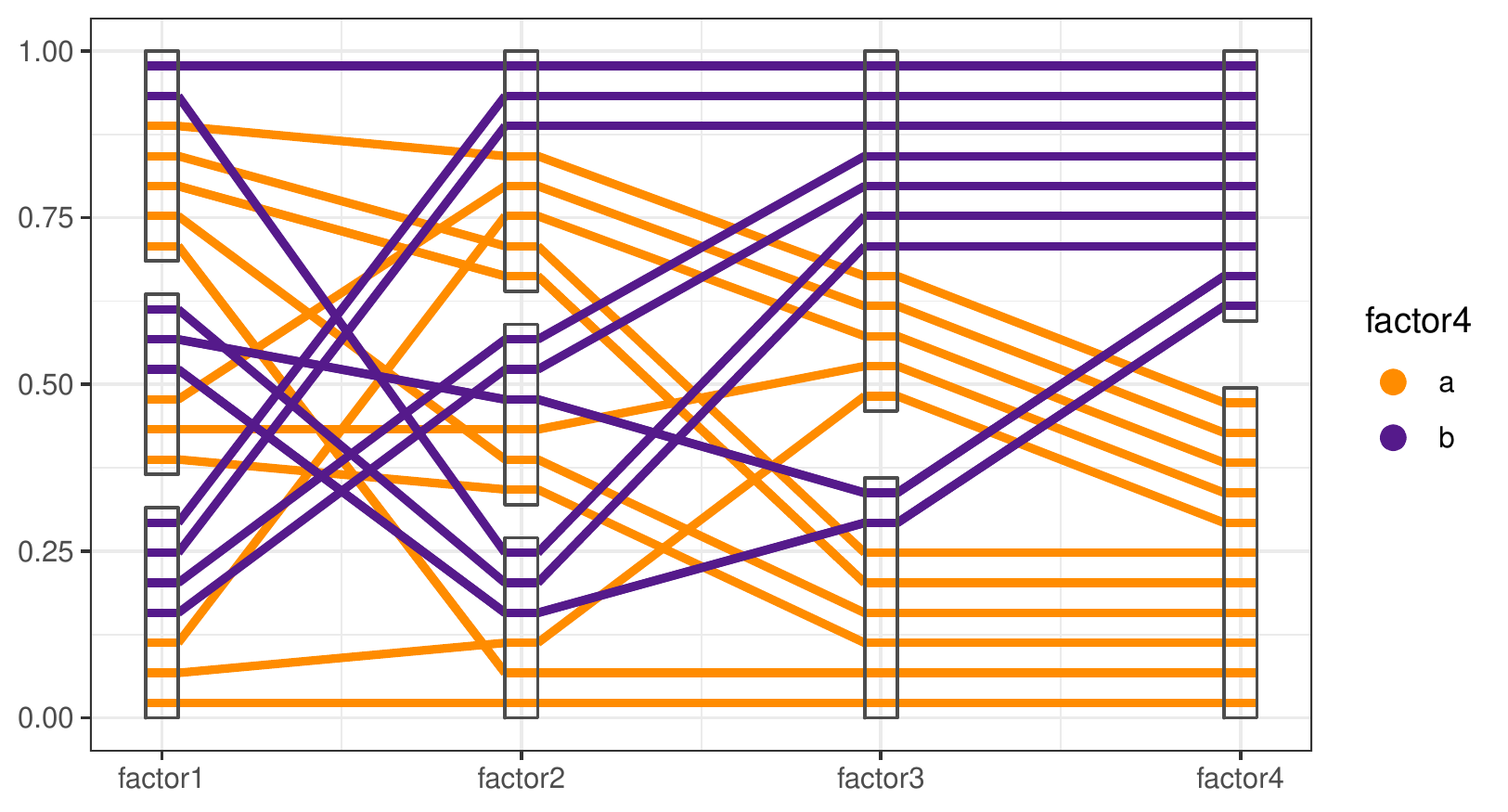} \caption{Factor block of three categorical variables. The order within each level of a categorial variable is determined by the joint distribution with all of the categorical variables on its right side.}\label{fig:factor-block}
\end{figure}

That means, that for the left most categorical variable, the positions
of the observations within each level are ordered according to their
corresponding positions in its adjacent categorical right-hand neighbor,
which itself is ordered by the position of observations of its
right-hand categorical neighbor. This is equivalent to an order given by
hierarchically sorting categorical variables from right to left. Any
remaining ties within a level are, as in the N-C case, resolved by the
order of values of the variable on the left. As a result, the parallel
coordinate plot appears to become gradually more ordered to the right,
as can be seen in \autoref{fig:factor-block}. This approach minimizes
the number of crossed lines.

\hypertarget{break-factor-blocks}{%
\subsection{Break Factor Blocks}\label{break-factor-blocks}}

\begin{figure}
\includegraphics[width=\linewidth]{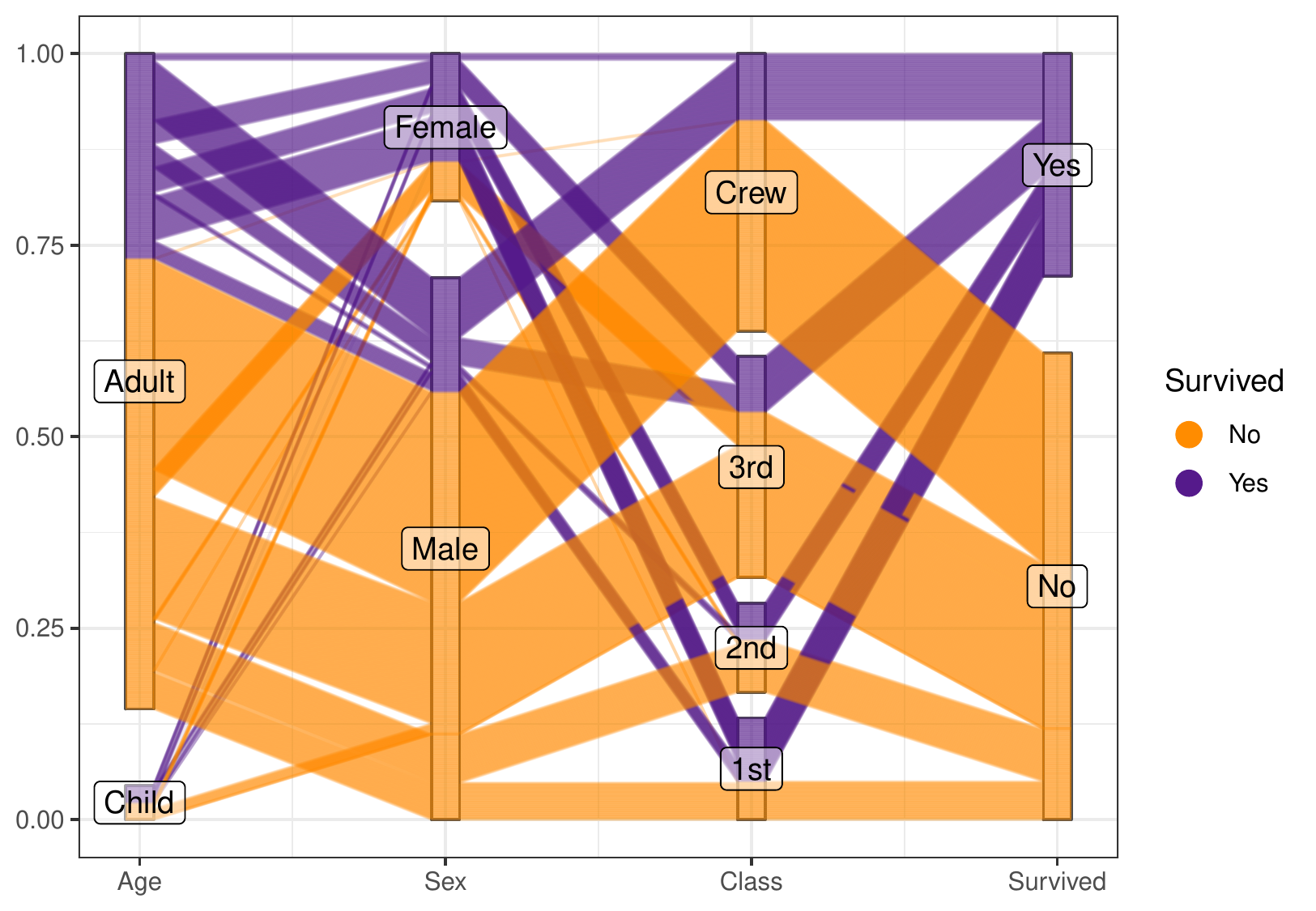} \caption{Fast accumulated combinations from right to left. Each line in the plot corresponds to one person on board the RMS Titanic. Lines are colored by survival. }\label{fig:titanic-full}
\end{figure}

While generalized parallel coordinate plots can now deal with multiple
categorical variables, we do have to pay a price in terms of complexity
by adding more and more categorical variables into the plot. As the
number of categorical variables in a factor block increases, the total
number of combinations in the corresponding joint distribution increases
multiplicatively. We can see the rapidly increasing number of
combinations in \autoref{fig:titanic-full}. This figure shows survival
on board the RMS Titanic by gender, age, and class \citep{dawson}. Only
for the right most section in the factor block can we see a simple
relationship between the eight possible combinations of survival and
passenger status/crew. The number of combinations rapidly devolves into
an incomprehensible mess moving from the ordered right hand side to the
left, as a result of utilizing the full joint distribution of the factor
combinations.

\begin{figure}
\includegraphics[width=\linewidth]{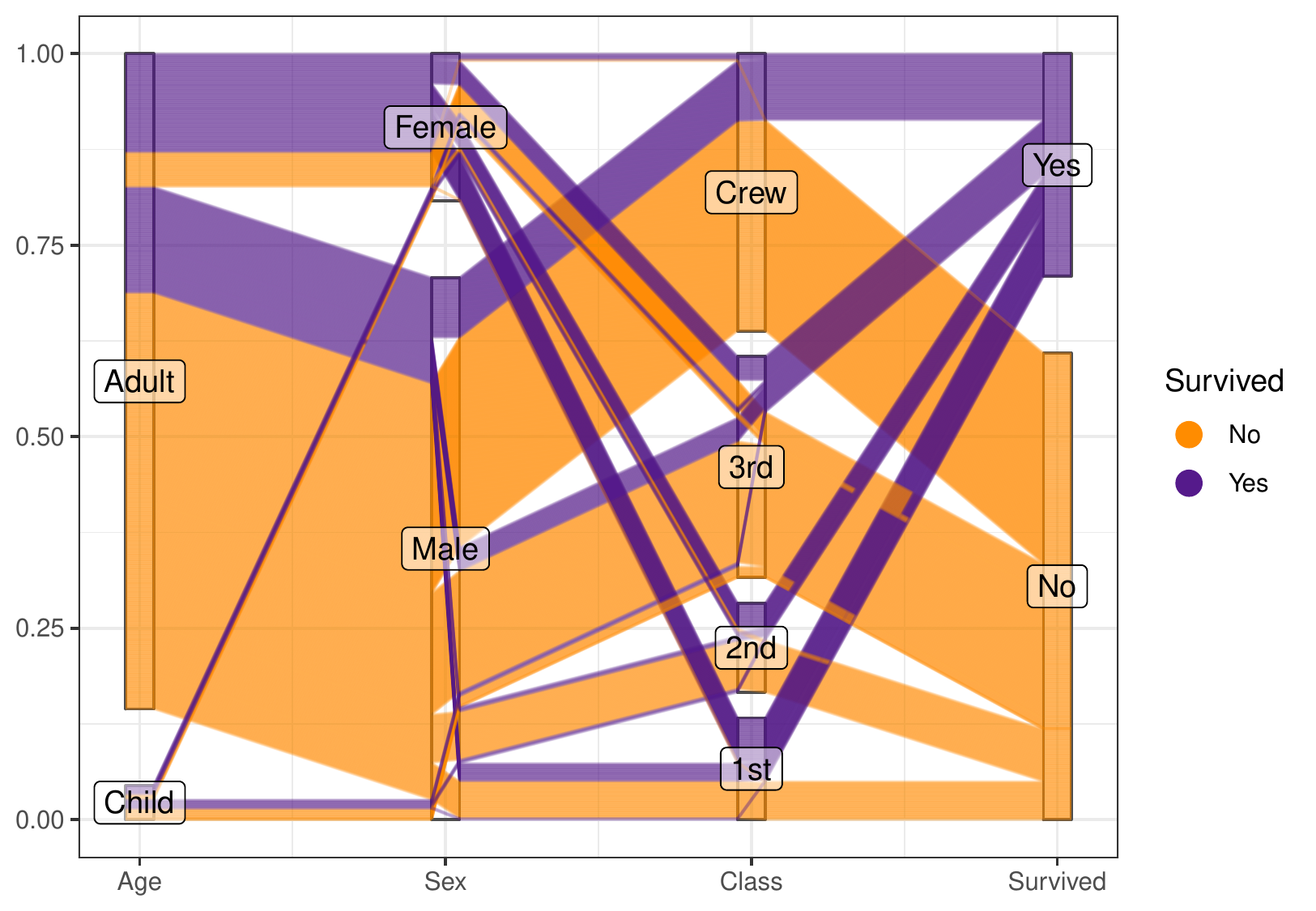} \caption{Same data and structure as the previous plot, with breakpoints inserted for the second and third variable. Relationships between adjacent variables are emphasized.}\label{fig:titanic-marg}
\end{figure}

In order to direct visual attention to the useful structure within the
data, rather than the increasing fragmentation of the joint distribution
of all of the displayed variables, we introduce a method for breaking
factor blocks into sub-blocks. The joint distribution of variables
within the sub-blocks is preserved, but between adjacent sub-blocks,
only the immediately relevant relationships are maintained. Visually,
this is manifested in a re-ordering of cases within a factor block,
shown by diagonal lines contained within the factor boxes. This hides
some of the complexity of the full joint distribution, but allows the
viewer to focus on the primary relationships of interest while
preserving most of the utility of the ordering described above.
\autoref{fig:titanic-marg} shows the same data as
\autoref{fig:titanic-full}, but has breakpoints inserted at each of the
interior variables; the re-ordering can be seen within the interior
blocks, and the resulting chart is cleaner and easier to read.

Computationally, the logic behind the factor block break points is as
follows: At each breakpoint, we arrange the right and left side of the
data separately, and then reconcile the two orderings within the
breakpoint box. This contains much of the visual clutter within the box
indicating the factor level, which reduces the visual impact of the
reordering significantly, but also makes it harder to track individual
observations across the plot.

\hypertarget{organized-over-plotting}{%
\subsection{Organized Over-plotting}\label{organized-over-plotting}}

One of the primary advantages of the generalized approach to categorical
variables described above is the ability to follow a single observation
throughout the plot. As the number of observations increase, this
becomes less feasible because with more polylines, there are more
intersections between polylines, exponentially increasing the effort
required to follow a polyline from one side of the plot to the other. To
reduce this tendency toward chaos, it is necessary to carefully control
the order in which lines are plotted to ensure that it is relatively
easy to follow a line (or group of polylines) across the plot. We have
developed three different methods to control this overplotting in order
to maintain visual order:

\begin{enumerate}
\def\labelenumi{\arabic{enumi}.}
\item
  Plot the smaller groups of lines on top of the larger groups,
\item
  Plot the larger groups of lines on top of the smaller groups of lines,
\item
  Use the hierarchical arrangement of factor variables to order the line
  plotting.
\end{enumerate}

The user can specify which ordering method should be used with an
additional parameter. When there are multiple factor blocks, or
breakpoints between factor blocks, it is necessary to reconcile plotting
order within axes as well, using the same type of logic. Note that we
have used two different overplotting methods in
\autoref{fig:titanic-full} and \autoref{fig:titanic-marg}. In
\autoref{fig:titanic-full} we plotted larger groups of lines on top of
smaller groups of lines, while for \autoref{fig:titanic-marg} we plotted
smaller groups of lines on top of the larger groups. The fact that we
can still see smaller lines in the first figure is due to the additional
use of \(\alpha\)-blending, i.e.~we are not drawing lines at full
saturation, but make them partially see-through. Usually, the
hierarchical arrangement produces the best plots.

\hypertarget{a-layered-approach-to-parallel-coordinate-plots}{%
\section{A Layered Approach to Parallel Coordinate
Plots}\label{a-layered-approach-to-parallel-coordinate-plots}}

Since its publication on CRAN in 2005, the R package {\bf ggplot2} has
seen a stellar rise in use with now tens of thousands of downloads a
day.

This success is due to the underlying conceptual framework. ggplot2 is
based on the Grammar of Graphics \citep{wilkinson:1999}, implemented and
adjusted for usability in R \citep{Wickham:2010}. This means that a plot
in ggplot2 is assembled descriptively as a set of layers. Each of these
layers consists of a functional mapping between the variables in the
data set and a component of the plot, such as an \(x\) or a \(y\) axis,
or plot specific aesthetics, such as the color or size of points.
Generally, layers have a single geometric representation (such as
e.g.~points, lines or rectangles).

What is unique about ggplot2 is that its implementation facilitates the
creation of extensions. Our package ggpcp is one such extension for
generalized parallel coordinate plots.

In accordance with the modular design principle of the tidyverse
\citep{tidyverse} we have developed a set of functions to deal with
separate aspects of generalized parallel coordinate plots.

Parallel coordinate plots are somewhat unique in that there is no
one-to-one mapping between a variable and an axis; instead, there are
arbitrarily many variables provided, and both the x and y positions are
calculated for each variable, and thus, each polyline. To accommodate
this complication, we have utilized the \texttt{vars} function used
throughout the tidyverse to allow the user to specify a set of variables
using the familiar syntax of the select helper functions found in the
tidyselect package. Interestingly, this also enables us to specify
variables multiple times in a plot, as shown in \autoref{fig:iris}.

The user primarily interacts with the \texttt{geom\_pcp} function, which
is constructed to handle the different aspects of parallel coordinate
plots through a consistent API. This function allows us to draw a set of
polylines as in a traditional parallel coordinate plot.

Consistent with the modular approach of ggplot2, this function only
draws lines, and the user must specify additional layers for additional
plot components, such as the boxes for categorical variables or text for
labels.

The code below generates \autoref{fig:NCN}. As can be seen, the layers
of ggpcp work seamlessly with functionality from the tidyverse and
elements of ggplot2.

\begin{verbatim}
set.seed(20191019)
iris %>%
  group_by(Species) %>%
  sample_n(10) %>%
  ggplot(aes(vars = vars(Sepal.Length, Species, 
                         Sepal.Width))) + 
  geom_pcp(boxwidth = 0.1, aes(color = Species), 
           size = 1.25) + 
  geom_pcp_box(boxwidth = 0.1) + 
  geom_pcp_label(boxwidth = 0.1) +
  theme_bw() +
  scale_colour_manual(
    values = c("purple4", "darkorange", "darkgreen")) 
\end{verbatim}

In the next section we will discuss some examples highlighting some more
sophisticated aspects of the plot.

\hypertarget{examples}{%
\section{Examples}\label{examples}}

\hypertarget{getting-a-second-third-and-seventh-opinion}{%
\subsection{Getting a second, third, \ldots{} and seventh
opinion}\label{getting-a-second-third-and-seventh-opinion}}

\begin{figure}
\includegraphics[width=\linewidth]{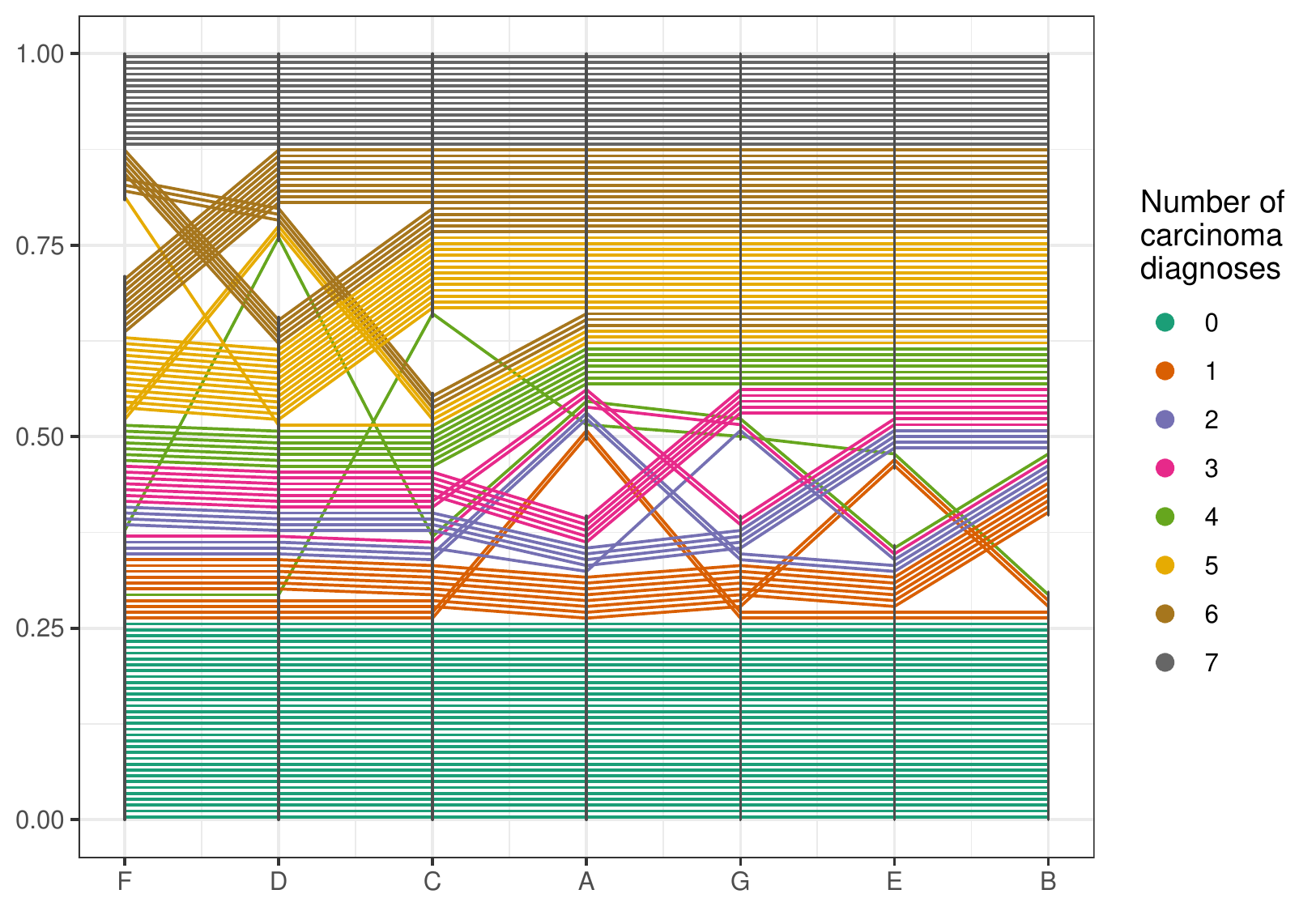} \caption{Pathologists' diagnoses of absence (no) or presence (yes) of carcinoma in the uterine cervix based on 118 slides. Each slide is shown by a polyline.}\label{fig:carcinoma}
\end{figure}

\autoref{fig:carcinoma} shows data from \citet{agresti} published as
part of the poLCA package \citep{polca}. Seven pathologists were asked
to assess the same 118 slides for the presence or absence of carcinoma
in the uterine cervix. Binary responses for each slide were recorded
(yes/no). Pathologists all agreed on about 25\% of slides, which they
considered to be carcinoma free, and a further 12.5\% of slides, which
were considered to show carcinoma by all pathologists.\\
For the remaining 62.5\% of slides there was some disagreement. However,
we see that this disagreement is not random. When pathologists are
ordered (by moving the corresponding axes) left to right from fewest
number of overall carcinoma diagnoses to highest number, we see that
generally for a slide more pathologists make a carcinoma diagnosis from
left to right.

\hypertarget{missing-migrants}{%
\subsection{Missing migrants}\label{missing-migrants}}

The Missing Migrants Project (\url{https://missingmigrants.iom.int/})
tracks incidents involving asylum seekers who have gone missing, were
injured or have died on the way to their destination. The project aims
to identify and track missing migrants and provide a reliable data
source for media, researchers and the general public. The Missing
Migrants Project started as a response to the tragedies of October 2013,
when at least 368 migrants died in shipwrecks near the Italian island of
Lampedusa. Here, we are exploring data from the three regions with the
highest number of incidents: North Africa, the Mediterranean and the
US-Mexico Border. In total we are considering 3273 incidents between Jan
1 2015 and Dec 31 2018.

\begin{figure}
\includegraphics[width=\linewidth]{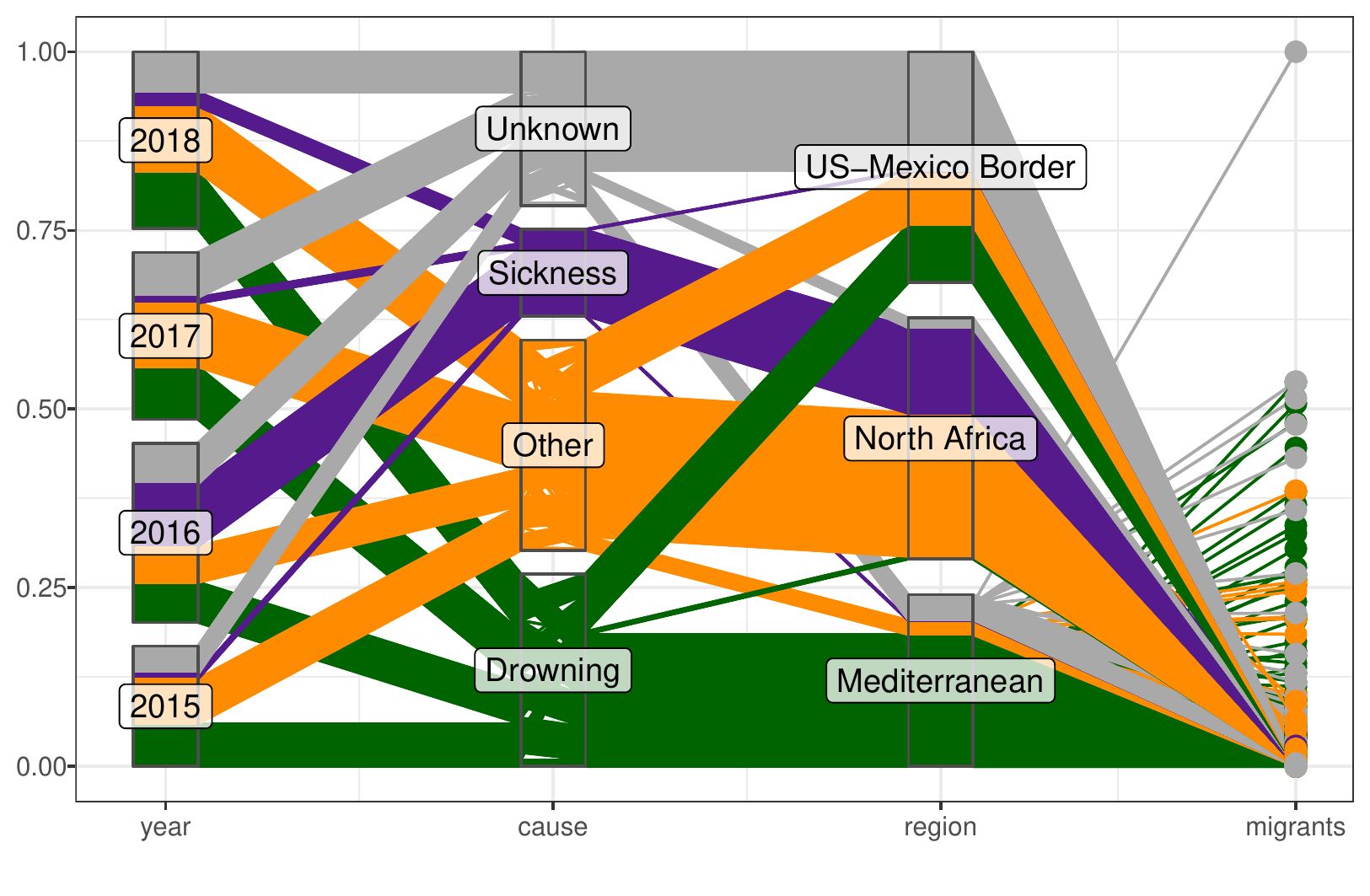} \caption{Missing migrants - generalized parallel coordinate plot of incidents involving migrants in three regions of the world. Each line  corresponds to one incident report. Lines are colored by cause of the incident. The variable on the right shows number of migrants involved in each incident.}\label{fig:mm2}
\end{figure}

Incidents are reported with several hundred classifications of causes.
Here, we are focusing on the top three: drowning, sickness, and unknown
and combine all other causes under `other'.

\autoref{fig:mm2} shows a generalized parallel coordinate plot of the
Missing Migrant Project data. Each line corresponds to one incident,
lines are colored by the cause of the incident. We can see that the
number of incidents in each of the three regions is similar, but in
terms of the number of people involved, the Mediterranean clearly
dominates the picture, with the worst incident estimated to have cost
the lives of more than 1000 people.

We see that in 2016 one leading cause of reported incidents were
sickness and diseases, mostly reported in North Africa. Further
investigation reveals that these are likely experienced by refugees from
Sudanese camps, where poor sanitation and complete lack of medical care
led to epidemic outbreaks of cholera, typhus and other preventable
diseases.

\begin{figure}
\includegraphics[width=\linewidth]{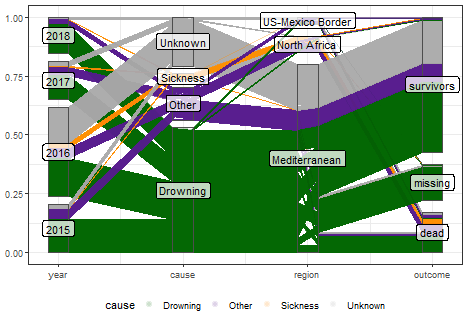} \caption{Missing migrants - each line corresponds to one migrant.}\label{fig:mm1}
\end{figure}

\autoref{fig:mm1} shows a generalized parallel coordinate plot of a
different aspect of the same data. Each line now corresponds to an
individual instead of a group involved in an incident. An outcome
variable is added to report on each individual's presumed status.

The number of migrants involved in incidents peaked in 2016 and numbers
have since decreased. Drowning is the leading cause of death for
migrants in the Mediterranean, but as can be seen in \autoref{fig:mm2}
there are a substantial number of drowning incidents along the US-Mexico
border.

Even though we saw in \autoref{fig:mm2} sickness as one of the leading
number of incident reports in 2016, fortunately the number of migrants
affected is relatively small.

\hypertarget{asa-data-expo-2006}{%
\subsection{ASA Data expo 2006}\label{asa-data-expo-2006}}

In this last example, we re-visit a data set that was used for the ASA
Data Expo in 2006. The \texttt{nasa} data, made available as part of the
\texttt{ggpcp} package provides an extension to the data provided in the
\texttt{GGally} package \citep{GGally}. It consists of monthly
measurements of several climate variables, such as cloud coverage,
temperature, pressure, and ozone values, captured on a 24x24 grid across
Central America between 1995 and 2000.

\begin{figure}

{\centering \includegraphics[width=.7\linewidth]{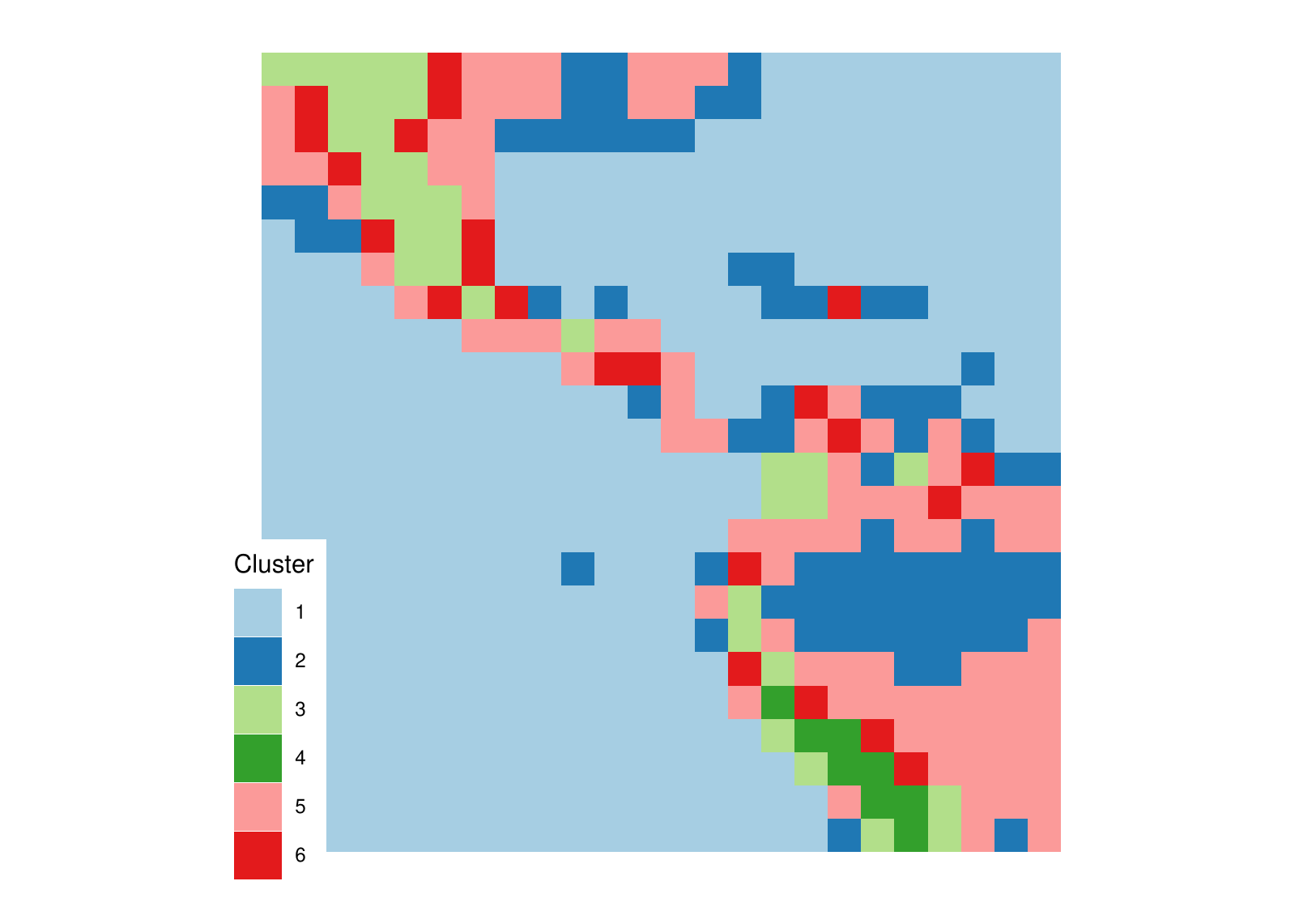} 

}

\caption{Tile plot of the (gridded) geographic area underlying the data. Each tile is colored by its cluster membership.}\label{fig:spatial}
\end{figure}

Using a hierarchical clustering (based on Ward's distance) of all
January and July measurements of all climate variables and the
elevation, we group locations into 6 clusters. The resulting cluster
membership can then be summarized visually. \autoref{fig:spatial} shows
a tile plot of the geography colored by cluster. We see that the
clusters have a very distinct geographic pattern.

\begin{figure}[h]
\includegraphics[width=\linewidth]{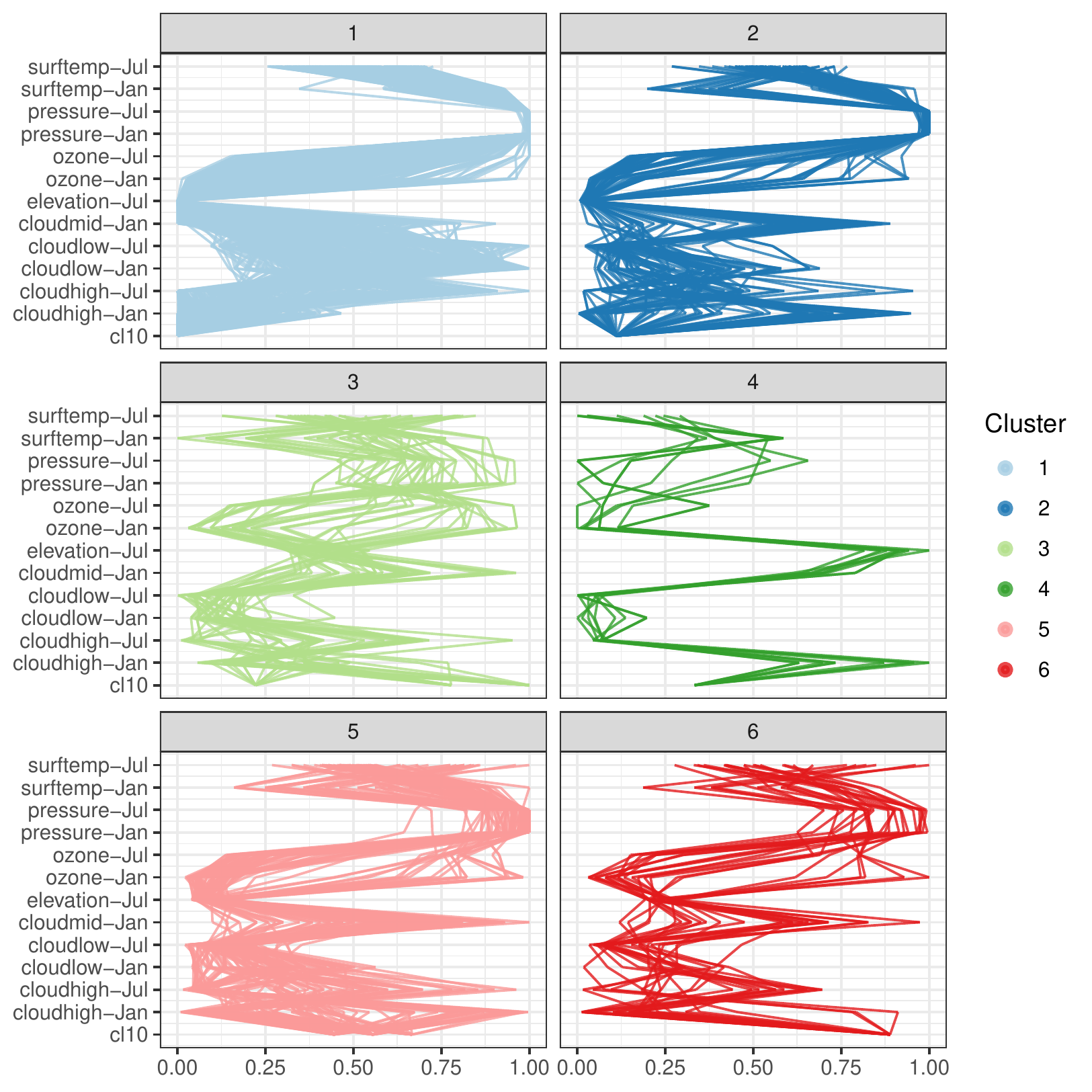} \caption{Overview of all variables involved in the clustering. }\label{fig:clusters}
\end{figure}

From the parallel coordinate plot in \autoref{fig:clusters} we see that
cloud coverage in low, medium and high altitude distinguishes quite
succinctly between some of the clusters. (Relative) temperatures in
January and July are very effective at separating between clusters in
the Southern and Northern hemisphere. The connection between the US gulf
coast line and the upper region of the Amazon (cluster 2) can probably
be explained by a relatively low elevation combined with similar
humidity levels.

A parallel coordinate plot allows us to visualize a part of the
dendrogram corresponding to the hierarchical clustering.

\begin{figure}
\includegraphics[width=\linewidth]{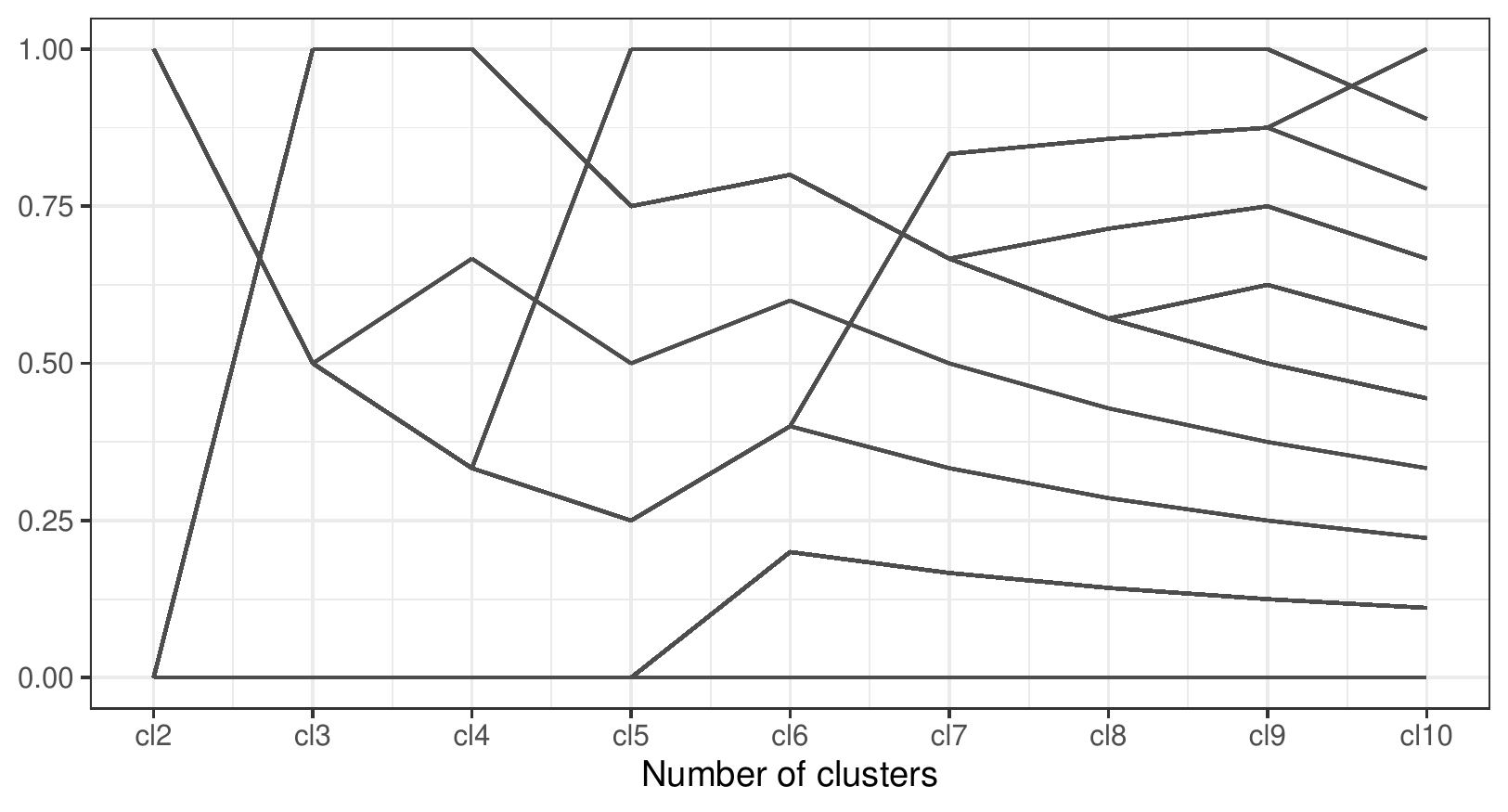} \caption{Dendrogram showing number of clusters at each step of the hierarchical process using the old-style parallel coordinate plot.}\label{fig:old}
\end{figure}

\begin{figure}
\includegraphics[width=\linewidth]{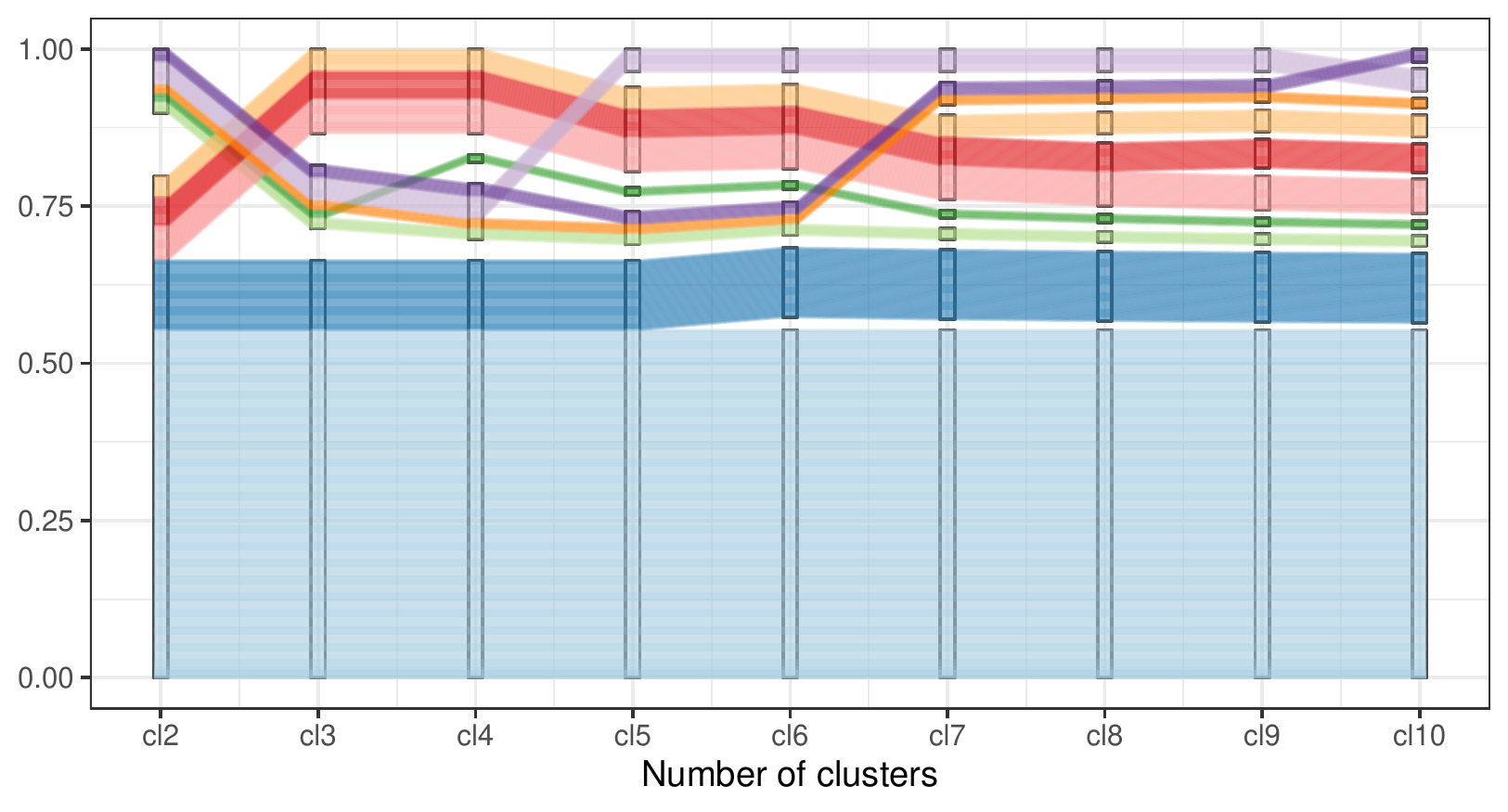} \caption{Same dendrogram as above using the much more informative generalized parallel coordinate plot.}\label{fig:clustergram}
\end{figure}

Using the generalized parallel coordinate plots we can visualize the
clustering process in plots similar to what Schonlau
\citep{clustergram1, clustergram2} coined the clustergram, see
\autoref{fig:old} and \autoref{fig:clustergram}.

Along the x-axis the number of clusters are plotted with one PCP axis
each, from two clusters (left) to 10 clusters (right most PCP axis).
Each line corresponds to one location, lines are colored by cluster
assignment in the ten-cluster solution. This essentially replicates the
dendrogram while providing information about the number of observations
in each cluster as well as the relationship between successive
clustering steps.

\clearpage

\hypertarget{discussion-and-further-work}{%
\section{Discussion and Further
Work}\label{discussion-and-further-work}}

The generalized parallel coordinate plot provides a visualization for a
mix of categorical and numeric variables, that incorporates existing
variants of numeric only and categorical variables only as (trivial)
special cases. Fundamental to this modified framework is the switch from
a frequency based representation of categorical variables to an
observation-based representation, which allows the viewer to track an
individual observation across the entire plot. While the observations
are drawn individually, visually the proximity of lines creates an
implicit grouping that characterizes the joint distribution of the data
and preserves the functionality of the original implementations of both
the categorical and the numeric variants of parallel sets/parallel
coordinate plots. Drawing lines for each individual might not be the
fastest strategy computationally, but it allows us to focus on the human
behind the data instead of aggregating the same information into a
faceless statistic.

In the implementation in {\bf ggpcp} we combined the extended PCP
framework with the powerful grammar of graphics implementation of
{\bf ggplot2}. The layer-based construction of generalized parallel
coordinate plots allows the user to explicitly describe each layer of
the plot in a familiar manner and therefore provides flexibility in
specifying and changing each aspect of the plot's appearance. As shown
in the example of migrant casualties, the tidyverse facilitates
transitioning between different levels of aggregation; the generalized
parallel coordinate plots tap into that power to provide visualizations
for different observational units. The ability to link between these
plots in an interactive manner might be achievable by additionally
leveraging the plotly framework \citep{plotly} and would further improve
upon the usefulness of generalized parallel coordinate plots

While we are quite convinced that the generalized parallel coordinate
plots are indeed an improvement over the traditional parallel coordinate
plot or parallel sets, we would like to confirm this in future user
studies.

While the \(\alpha\)-blending of the lines mitigates some of the effect
of the sine illusion, the exact magnitude of the mitigation heavily
depends on the choice of \(\alpha\) and the data set. Future versions of
the package might implement the idea of the common angle plot, therefore
adjusting for the effect by forcing all lines to appear under the same
angle. This might also help with reducing the visual complexity.

\clearpage

\bibliographystyle{agsm}
\bibliography{bibliography.bib}

\end{document}